\documentclass[aps,prd,twocolumn,10pt]{revtex4-2}

\usepackage{amsmath,amssymb}
\usepackage{booktabs}
\usepackage{graphicx}
\usepackage[colorlinks=true,allcolors=blue]{hyperref}
\usepackage[normalem]{ulem}
\usepackage{xcolor}

\usepackage{xcolor}
\usepackage{soul}

\setstcolor{red}  
\begin{document}

\title{Probing Velocity-Dependent WIMP Dark Matter Annihilation in Large-Scale Structure}

\author{Wen-Qing Guo}
\email{wetingguo@gmail.com}
\affiliation{Department of Physics, Stellenbosch University, Matieland 7602, South Africa}

\begin{abstract}
The unresolved $\gamma$-ray background (UGRB) provides a promising avenue for probing the cosmological annihilation of weakly interacting massive particle (WIMP) dark matter through its cross-correlation with large-scale structure (LSS). While most previous studies have assumed a velocity-independent ($s$-wave) annihilation cross section, many well-motivated particle physics models predict annihilation rates that depend on the relative velocity of dark matter particles. In this work, we develop a generalized halo-model framework to investigate the cosmological signatures of velocity-dependent WIMP annihilation. Parameterizing the velocity-weighted annihilation cross section as $\sigma v_{\rm rel}\propto v_{\rm rel}^{n}$, we consider four representative cases corresponding to Sommerfeld-like ($n=-1$), $s$-wave ($n=0$), $p$-wave ($n=2$), and $d$-wave ($n=4$) annihilation. We show that velocity dependence modifies not only the overall annihilation emission but also the halo population dominating the signal, leading to distinct effective halo biases, redshift distributions, and angular clustering properties. As a result, different annihilation mechanisms leave characteristic signatures in the UGRB--LSS cross-correlation that cannot be captured by the overall $\gamma$-ray intensity alone. As an illustration of the formalism, we perform a statistical forecast for the cross-correlation between the \textit{Fermi} Large Area Telescope (LAT) UGRB and the Two Micron All Sky Survey Redshift Survey (2MRS) galaxy catalog and derive the projected sensitivities to the velocity-independent annihilation coefficient for these four annihilation scenarios. Our results demonstrate that cosmological cross-correlations provide a powerful and complementary probe of velocity-dependent dark matter annihilation, offering a new avenue for constraining the particle physics of dark matter through its imprints on large-scale structure.
\end{abstract}

\maketitle

\section{Introduction}
Dark matter (DM) constitutes approximately $85\%$ of the matter content of the Universe and plays a pivotal role in the formation of cosmic structures, yet its particle nature remains one of the most profound open questions in modern physics \cite{Planck:2018vyg}. Among the vast landscape of candidates, weakly interacting massive particles (WIMPs) stand out as one of the most compelling scenarios, as they naturally reproduce the observed relic abundance via the thermal freeze-out mechanism \cite{Roszkowski:2017nbc, Arcadi:2017kky}. In this paradigm, WIMP annihilations into Standard Model particles produce stable secondary messengers, including $\gamma$ rays, neutrinos, and cosmic rays, offering unique opportunities for indirect detection \cite{Hooper:2010mq, Huang:2011xr, Funk:2013gxa, Conrad:2014tla, Fermi-LAT:2015att, Klasen:2015uma, Leane:2020liq, Basu:2021zfg, Guo:2022rqq, DiMauro:2023qat}.

On cosmological scales, the cumulative annihilation of WIMPs within dark matter halos gives rise to a diffuse $\gamma$-ray background. Since the majority of these halos remain unresolved by current $\gamma$-ray telescopes, their collective emission contributes appreciably to the unresolved $\gamma$-ray background (UGRB), which has been precisely measured across a broad energy range by the \textit{Fermi} Large Area Telescope (LAT) \cite{Fermi-LAT:2012pez, Fermi-LAT:2014ryh, Fermi-LAT:2018udj, Fornasa:2016ohl}. Over the past decade, measurements of the UGRB intensity, energy spectrum, and angular anisotropies have been extensively leveraged to probe WIMP annihilation or decay signatures. Since dark matter halos trace the underlying large-scale matter distribution, the $\gamma$-ray emission produced by WIMP annihilations within these halos is naturally expected to correlate spatially with tracers of the large-scale structure (LSS). In particular, cross-correlation analyses between the UGRB and various LSS tracers, including cosmic shear \cite{Camera:2012cj, Camera:2014rja, DES:2025ulp, Zhang:2026ysp}, galaxy catalogs \cite{Regis:2015zka, Ammazzalorso:2018evf, Zhou:2024cld, Pinetti:2025hgd, Rubiola:2025scy}, galaxy clusters \cite{Tan:2019gmb}, and cosmic voids \cite{Arcari:2022zul}, provide a powerful strategy to suppress uncorrelated astrophysical backgrounds and enhance sensitivity to faint dark-matter-induced signals.

Despite the effectiveness of these cross-correlation studies, most existing analyses adopt the simplifying assumption of a velocity-independent ($s$-wave) annihilation cross section. In reality, however, the WIMP annihilation rate can exhibit a nontrivial dependence on the relative velocity of the annihilating particles. In general, the velocity-weighted cross section admits a partial-wave expansion, with the leading-order terms corresponding to $s$-, $p$-, and $d$-wave channels, characterized by $\sigma v_{\rm rel} \propto v_{\rm rel}^{0}$, $v_{\rm rel}^{2}$, and $v_{\rm rel}^{4}$, respectively \cite{Ferrer:2013cla, Boddy:2017vpe, Zhao:2017dln, Petac:2018gue, Boddy:2018ike, Boddy:2018qur, Kiriu:2022bjq, Blanchette:2022hir}. Moreover, non-perturbative effects such as the Sommerfeld enhancement can introduce an inverse scaling $\sigma v_{\rm rel} \propto v_{\rm rel}^{-1}$ in the low-velocity regime, potentially boosting the annihilation rate by orders of magnitude \cite{Arkani-Hamed:2008hhe, Robertson:2009bh, Slatyer:2009vg, Feng:2010zp, Beneke:2022rjv}. Therefore, the velocity-dependent annihilation signal depends not only on the dark matter density field but also on the velocity distribution within individual halos. For instance, $p$- and $d$-wave annihilations preferentially probe high-velocity environments, whereas Sommerfeld-enhanced interactions are most prominent in low-velocity systems. Consequently, halos of different masses, concentrations, and dynamical states contribute differently to the total annihilation signal depending on the underlying velocity scaling. These considerations suggest that the velocity dependence of WIMP annihilation can leave distinct imprints on the spatial and redshift distributions of the annihilation signal embedded in the UGRB \cite{Campbell:2010xc, Campbell:2011kf, Boddy:2019wfg}. This raises the intriguing possibility that cross-correlating the UGRB with LSS tracers can serve as a powerful probe to distinguish between different partial-wave scenarios.

In this work, we investigate the potential of UGRB-LSS cross-correlations to probe the velocity dependence of WIMP annihilation. We adopt a unified parameterization of the velocity-weighted annihilation cross section, $ \sigma v_{\rm rel} \propto v_{\rm rel}^{n}$, and examine four representative cases corresponding to Sommerfeld-like ($n=-1$; {Coulomb limit with a massless mediator}), $s$-wave ($n=0$), $p$-wave ($n=2$), and $d$-wave ($n=4$) annihilation. Within this framework, we compute the dark-matter-induced contribution to the cross angular power spectrum (APS) between the UGRB measured by \textit{Fermi}-LAT and the Two Micron All Sky Survey Redshift Survey (2MRS) galaxy catalog. We systematically study how different velocity dependencies modify the spatial clustering and redshift distribution of the signal, and perform a sensitivity forecast to assess the ability of current-generation instruments to constrain the velocity-independent annihilation coefficient for several annihilation scenarios. Our results demonstrate that the UGRB-LSS cross-correlation provides meaningful and complementary information on the velocity dependence of dark matter annihilation, offering a largely unexplored avenue beyond conventional searches.

This paper is organized as follows. In Sec.~\ref{Sec:Theore}, we outline the theoretical framework for computing the angular power spectrum. Our forecast methodology and results are presented and discussed in Sec.~\ref{Sec:ResultDiscussion}. We summarize our conclusions in Sec.~\ref{Sec:Conclusion}.

Throughout this paper, {velocities are normalized by the speed of light $c$, and} we adopt a spatially flat $\Lambda$CDM cosmology with the \textit{Planck} 2018 cosmological parameters: $H_{0}=67.32~{\rm km\,s^{-1}Mpc^{-1}}$, $\Omega_{\rm c}h^{2}=0.1201$, and $\Omega_{\rm b}h^{2}=0.02238$ \cite{Planck:2018vyg}.

\begin{figure*}[t]
    \centering
    \includegraphics[width=0.48\linewidth]{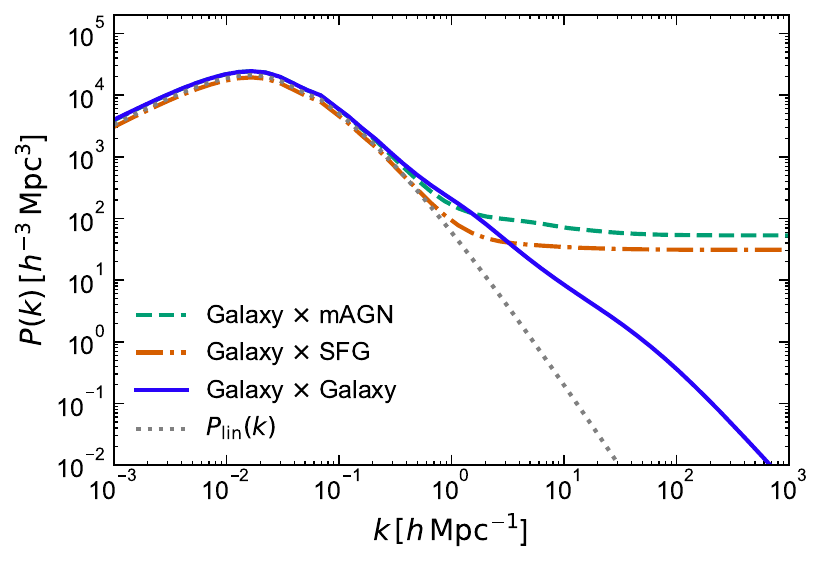}
    \includegraphics[width=0.48\linewidth]{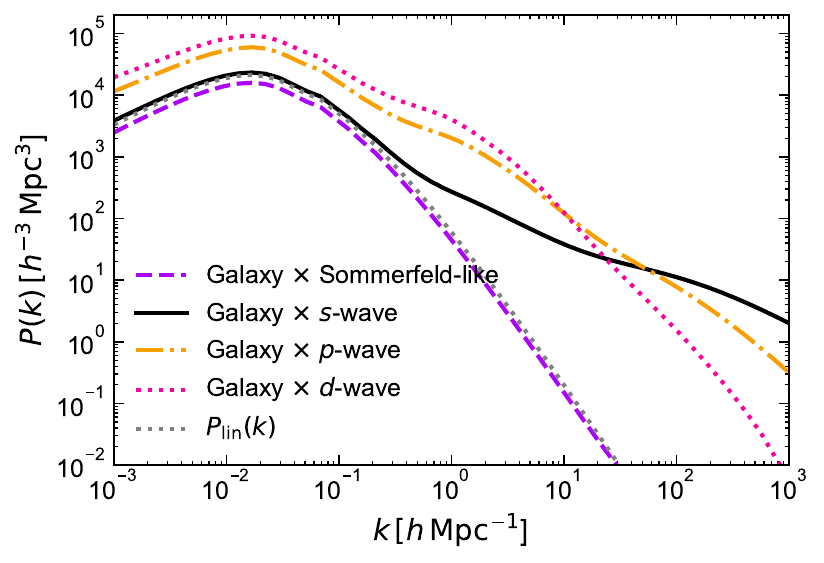}
    \caption{Three-dimensional power spectra at $z=0.05$. \textit{Left}: The solid blue line shows the galaxy auto-power spectrum, while the dashed green and dash-dotted brown lines show the galaxy--$\gamma$-ray cross-power spectra for unresolved mAGNs and SFGs, respectively, in the $(4.8,\,8.3)\,\mathrm{GeV}$ energy bin. \textit{Right}: The galaxy--$\gamma$-ray cross-power spectra for the Sommerfeld-like (dashed purple), $s$-wave (solid black), $p$-wave (dash-dotted orange), and $d$-wave (dotted pink) annihilation scenarios. The gray dotted line in both panels shows the linear matter power spectrum at the same redshift.}
    \label{fig:ThreeD_PS}
\end{figure*}

\section{The cross-correlation signal}\label{Sec:Theore}
\subsection{Angular power spectrum}
We quantify the cross-correlation between two cosmological observables $i$ and $j$ via their cross angular power spectrum. Under the Limber approximation, the cross APS $C_{\ell}^{ij}$ is given by \cite{Fornengo:2013rga}
\begin{equation}\label{eq:APS}
C_{\ell}^{ij} = \int \frac{\mathrm{d}\chi}{\chi^{2}} W_i(\chi) W_j(\chi) P_{ij}\!\left(z, k=\frac{\ell+1/2}{\chi}\right),
\end{equation}
where $\ell$ denotes the angular multipole, $\chi(z)$ is the comoving distance with $\mathrm{d}\chi = c\,\mathrm{d}z/H(z)$, and $W(\chi)$ is the window function encoding the redshift distribution of each observable. The Hubble expansion rate in a flat $\Lambda$CDM cosmology is
\begin{equation}
H(z) = H_0 \sqrt{\Omega_{\rm m}(1+z)^3 + \Omega_\Lambda},
\end{equation}
with $\Omega_\Lambda = 1 - \Omega_{\rm m}$ and $\Omega_{\rm m} = 0.3158$. In what follows, we use $\chi$ and $z$ interchangeably.

The term $P_{ij}(z,k)$ in Eq.~\eqref{eq:APS} is the three-dimensional cross-power spectrum of the fluctuation fields of the two observables. Adopting the halo-model framework \cite{Cooray:2002dia}, we decompose $P_{ij}(z,k)$ into one-halo (1h) and two-halo (2h) terms
\begin{equation}
P_{ij}(z,k) = P_{ij}^{1\mathrm{h}}(z,k) + P_{ij}^{2\mathrm{h}}(z,k).
\end{equation}
The one-halo term captures correlations between tracers within the same halo, which dominate on small scales, whereas the two-halo term describes correlations between tracers residing in distinct halos and governs the large-scale behavior. These contributions are expressed as
\begin{align}
P^{1\mathrm{h}}_{ij}(z,k) &= \int \mathrm{d}M \frac{\mathrm{d}n}{\mathrm{d}M}(z,M) \frac{\tilde g_i^{*}(k|z,M)}{\langle g_i\rangle(z)} \frac{\tilde g_j(k|z,M)}{\langle g_j\rangle(z)}, \\
P^{2\mathrm{h}}_{ij}(z,k) &= \left[ \int \mathrm{d}M_1 \frac{\mathrm{d}n}{\mathrm{d}M_1} b_{\rm h}(z,M_1) \frac{\tilde g_i^{*}(k|z,M_1)}{\langle g_i\rangle(z)} \right] \nonumber \\
&\quad \times \left[ \int \mathrm{d}M_2 \frac{\mathrm{d}n}{\mathrm{d}M_2} b_{\rm h}(z,M_2) \frac{\tilde g_j(k|z,M_2)}{\langle g_j\rangle(z)} \right] P_{\rm lin}(z,k),
\end{align}
where $M$ is the halo mass, and the integration is performed over the mass range $M_{\rm min} = 10^{-6}\,\mathrm{M_\odot}$ to $M_{\rm max} = 10^{18}\,\mathrm{M_\odot}$. The halo mass function $\mathrm{d}n/\mathrm{d}M$ and the halo bias $b_{\rm h}$ are modeled using the Sheth-Tormen prescriptions \cite{Sheth:1999su, Sheth:1999mn}. The function $\tilde g(k|z,M)$ is the Fourier transform of the spatial profile $g(r|z,M)$ of a given tracer within a halo of mass $M$, with the asterisk denoting complex conjugation. The linear matter power spectrum $P_{\rm lin}(z,k)$ is obtained from the public code {\tt CAMB}\footnote{\url{https://camb.readthedocs.io/en/latest/}}.

In this work, we specifically examine the cross-correlation between the 2MRS galaxy catalog and the UGRB as measured by the \textit{Fermi}-LAT. The corresponding window functions, spatial profiles, and halo-model ingredients for these two tracers are described in the subsections below.

\subsection{Galaxy catalog}
As our large-scale structure tracer, we adopt the Two Micron All Sky Survey Redshift Survey (2MRS) \cite{2012ApJS..199...26H}, which provides a spectroscopic galaxy catalog in the local Universe with $N_{\rm g} = 43,182$ galaxies out to $z \simeq 0.1$ and an effective sky fraction $f_{\rm 2MRS} = 0.877$ after masking \cite{2006AJ....131.1163S, Ando:2017wff}. The normalized redshift distribution of 2MRS galaxies follows the empirical form \cite{Ando:2017wff}
\begin{equation}
\frac{\mathrm{d}N_{\rm 2MRS}}{\mathrm{d}z} = \frac{\beta}{z_0\Gamma[(m+1)/\beta]} \left( \frac{z}{z_0} \right)^m \exp \left[ - \left( \frac{z}{z_0} \right)^\beta \right],
\end{equation}
with best-fit parameters $\beta = 1.64$, $z_0 = 0.0266$, and $m = 1.31$. The corresponding window function entering Eq.~\eqref{eq:APS} is $W_{\rm g}(\chi) = (\mathrm{d}N_{\rm 2MRS}/\mathrm{d}z)\,(H(z)/c)$. 

To connect galaxies to the underlying dark matter distribution, we adopt the Halo Occupation Distribution (HOD) framework \cite{Zheng:2004id, Ando:2014aoa, Ammazzalorso:2018evf}. In this prescription, each halo of mass $M$ hosts a central galaxy at the halo center and a population of satellite galaxies tracing the smooth halo profile. Therefore the galaxy number density profile within a halo is given by
\begin{equation}
\begin{aligned}
g_{\rm g}(\mathbf{x}-\mathbf{x}'|z,M)=&\,N_{\rm cen}(M)\delta_{\rm D}^{(3)}(\mathbf{x}-\mathbf{x}')\\
&+N_{\rm sat}(M)\frac{\rho_{\rm DM}(\mathbf{x}-\mathbf{x}'|z,M)}{M},
\end{aligned}
\end{equation}
where $\delta_{\rm D}^{(3)}$ represents the three-dimensional Dirac delta function. We model the halo profile using the standard Navarro-Frenk-White (NFW) profile
\begin{equation}
\rho_{\rm DM}(r) = \frac{\rho_{\rm s}}{(r/r_{\rm s})(1+r/r_{\rm s})^2},
\end{equation}
{where the scale radius is related to the virial radius $R_{\rm vir}(z, M)$ by $r_{\rm s} = R_{\rm vir}/c_{\rm vir}$. The concentration parameter $c_{\rm vir}(z,M)$ is computed using Eq. (18) of Ref.~\cite{Coe:2010xg}, which requires $c_{200}$ as input. For the latter, we adopt the $c_{200}$--$M$ relation of Ref.~\cite{Correa:2015dva} calibrated for $z\le4$ under the \textit{Planck} cosmology.} The mean occupation numbers for central and satellite galaxies are parameterized as 
\begin{align}
N_{\rm cen}(M) &= \frac{1}{2} \left[ 1 + \mathrm{erf} \left( \frac{\log_{10}M - \log_{10}M_{\rm min}}{\sigma_{\log_{10} M}} \right) \right], \\
N_{\rm sat}(M) &= \begin{cases} \left( \frac{M-M_0}{M_1} \right)^\alpha, & M \ge M_0, \\ 0, & M < M_0, \end{cases}
\end{align}
with parameters for the 2MRS catalog taken from Ref.~\cite{Ando:2014aoa}, set to $\log_{10}(M_{\rm min}/\mathrm{M}_\odot) = 11.68$, $\sigma_{\log_{10}M} = 0.15$, $\log_{10}(M_0/\mathrm{M}_\odot) = 11.86$, $\log_{10}(M_1/\mathrm{M}_\odot) = 13.0$, and $\alpha = 1.02$. The mean galaxy number density is then
\begin{equation}
\langle g_{\rm g}\rangle(z) = \int \mathrm{d}M \frac{\mathrm{d}n}{\mathrm{d}M}(z,M) \int \mathrm{d}^3\mathbf{x}\, g_{\rm g}(\mathbf{x}-\mathbf{x}'|z,M),
\end{equation}
and the Fourier transform of the galaxy profile reads
\begin{equation}
\tilde g_{\rm g}(k|z,M) = N_{\rm cen}(M) + \frac{N_{\rm sat}(M)}{M}\,\tilde\rho_{\rm DM}(k|z,M),
\end{equation}
where $\tilde\rho_{\rm DM}$ is the Fourier transform of the NFW profile. For completeness, we define the effective linear galaxy bias as
\begin{equation}
b_{\rm g}(z) = \int \mathrm{d}M \, \frac{\mathrm{d}n}{\mathrm{d}M} \, b_{\rm h}(z,M) \, \frac{N_{\rm cen}(M) + N_{\rm sat}(M)}{\langle g_{\rm g}\rangle(z)}.
\end{equation}

\begin{table*}[t]
\centering
\setlength{\tabcolsep}{10pt}
\caption{Adopted \textit{Fermi}-LAT observational specifications used in this work, following Ref.~\cite{Fermi-LAT:2018udj}. For each energy bin, we list the energy range, the photon Poisson noise $C_{\rm N}^{\gamma}$, and the effective sky fraction after masking the Galactic plane and resolved $\gamma$-ray sources.}
\label{tab:FermiLAT_specifications}
\begin{tabular}{lcccc}
\toprule
Energy Bin &
$E_{\rm min}$ [GeV] &
$E_{\rm max}$ [GeV] &
$C_{\rm N}^{\gamma}$ [cm$^{-4}$\,s$^{-2}$\,sr$^{-1}$] &
$f_{\rm Fermi}$ \\
\midrule
1  & 0.5   & 1.0    & $1.056\times10^{-17}$ & 0.199 \\
2  & 1.0   & 1.7    & $3.548\times10^{-18}$ & 0.250 \\
3  & 1.7   & 2.8    & $1.375\times10^{-18}$ & 0.443 \\
4  & 2.8   & 4.8    & $8.324\times10^{-19}$ & 0.511 \\
5  & 4.8   & 8.3    & $3.904\times10^{-19}$ & 0.564 \\
6  & 8.3   & 14.5   & $1.768\times10^{-19}$ & 0.586 \\
7  & 14.5  & 22.9   & $6.899\times10^{-20}$ & 0.586 \\
8  & 22.9  & 39.8   & $3.895\times10^{-20}$ & 0.586 \\
9  & 39.8  & 69.2   & $1.576\times10^{-20}$ & 0.586 \\
10 & 69.2  & 120.2  & $6.205\times10^{-21}$ & 0.586 \\
11 & 120.2 & 331.1  & $3.287\times10^{-21}$ & 0.586 \\
12 & 331.1 & 1000.0 & $5.094\times10^{-22}$ & 0.586 \\
\bottomrule
\end{tabular}
\end{table*}

The resulting galaxy auto-power spectrum at $z=0.05$ is shown in the left panel of Fig.~\ref{fig:ThreeD_PS} (solid blue curve), with the linear matter power spectrum overlaid for reference (dotted gray line).

\subsection{$\gamma$-ray emission}
To forecast the sensitivity of the UGRB-galaxy cross-correlation to the WIMP-annihilation-induced signature, we adopt the observational specifications of the \textit{Fermi}-LAT from Ref.~\cite{Fermi-LAT:2018udj}, including energy binning, photon Poisson noise, and sky fractions, which are summarized in Table~\ref{tab:FermiLAT_specifications}.

The UGRB is generally believed to be dominated by unresolved astrophysical source populations, including BL Lacertae objects (BL Lac), flat spectrum radio quasars (FSRQ), misaligned active galactic nuclei (mAGN), and star-forming galaxies (SFGs) \cite{Ando:2009nk, Xia:2011ax, DiMauro:2014wha, Xia:2015wka, Hooper:2016gjy, Cuoco:2017bpv}. Nevertheless, the relative contributions of these populations remain uncertain, leaving room for additional components, including a potential contribution from dark matter annihilation \cite{Ando:2005xg, Fornengo:2013rga, Fornasa:2016ohl, Pinetti:2019ztr, Zhou:2024cld}.
We then model the UGRB as the sum of two components
\begin{equation}
I_{\gamma}^{\rm tot} = I_{\gamma}^{\star} + I_{\gamma}^{\rm WIMP},
\end{equation}
where $I_{\gamma}^{\star}$ denotes the cumulative emission from unresolved astrophysical sources, and $I_{\gamma}^{\rm WIMP}$ is the dark matter annihilation signal. For the astrophysical baseline, we include the two dominant low-redshift ($z\leq 0.1$) source populations, namely misaligned active galactic nuclei (mAGNs) and star-forming galaxies (SFGs) \cite{Pinetti:2019ztr, Blanco:2021icw, Zhou:2024cld}. The full modeling of the astrophysical components is presented in detail in our previous work \cite{Guo:2026ezl}. Here we briefly recapitulate the essential ingredients for completeness.

For a generic unresolved source population, the ensemble-averaged $\gamma$-ray luminosity density is
\begin{equation}
\langle g_{\star}\rangle(z) = \int_{L_{\rm min}}^{L_{\rm th}(z)} \mathrm{d}L_{\gamma} \, \phi_{\gamma}(z, L_{\gamma}),
\end{equation}
where $\phi_{\gamma}(z, L_{\gamma}) = \mathrm{d}^2 n / (\mathrm{d}V \, \mathrm{d}\ln L_{\gamma})$ is the $\gamma$-ray luminosity function (GLF), and $L_{\rm th}(z)$ is the redshift-dependent detection threshold of the instrument (derived in detail in Ref. \cite{Guo:2026ezl}). The window function of the $\gamma$-ray emission from unresolved astrophysical sources at observed energy $E_{\gamma}$ is given by \cite{Ando:2013ff}
\begin{equation}
\begin{aligned}
W_\star(E_{\gamma},\chi) = & \frac{d_{\rm L}^2(z)}{(1+z)^2} \int_{L_{\rm min}}^{L_{\rm th}(z)} \mathrm{d}\ln L_{\gamma} \, \phi_{\gamma}(z, L_{\gamma}) \\
& \times \frac{\mathrm{d}F}{\mathrm{d}E_{\gamma}}(E_{\gamma},z,L_{\gamma}) e^{-\tau([1+z]E_{\gamma})},
\end{aligned}
\end{equation}
where $d_{\rm L}(z)$ is the luminosity distance, and the exponential factor accounts for attenuation by the diffuse Extragalactic Background Light (EBL) with optical depth $\tau$ from Ref.~\cite{Razzaque:2008te}. Assuming power-law spectra with index $\Gamma$, the observed differential flux is \cite{Guo:2026ezl}
\begin{equation}
\frac{\mathrm{d}F}{\mathrm{d}E_{\gamma}} = \frac{L_{\gamma}}{4\pi d_{\rm L}^2} \frac{2-\Gamma}{E_{\gamma,\max}^{2-\Gamma} - E_{\gamma,\min}^{2-\Gamma}} E_{\gamma}^{-\Gamma},
\end{equation}
with $E_{\gamma,\min}=0.1~\mathrm{GeV}$ and $E_{\gamma,\max}=1~\mathrm{TeV}$ \cite{2003A&A...402..443I, Fermi-LAT:2012nqz, Madejski:2016oqg, Kornecki:2025tej}. The effective bias of the unresolved population is
\begin{equation}
b_{\star}(z) = \int_{L_{\rm min}}^{L_{\rm th}(z)} \mathrm{d}L_{\gamma} \, \phi_{\gamma}(z,L_{\gamma}) \, \frac{b_{\rm h}(z,M(L_{\gamma}))}{\langle g_{\star}\rangle(z)}.
\end{equation}
Evaluating this integral relies on the connection between the host halo mass $M$ and the intrinsic $\gamma$-ray luminosity $L_{\gamma}$ of the source, as detailed in the following subsections.

\subsubsection{Misaligned AGN}

\begin{figure*}[t]
    \centering
    \includegraphics[width=0.48\linewidth]{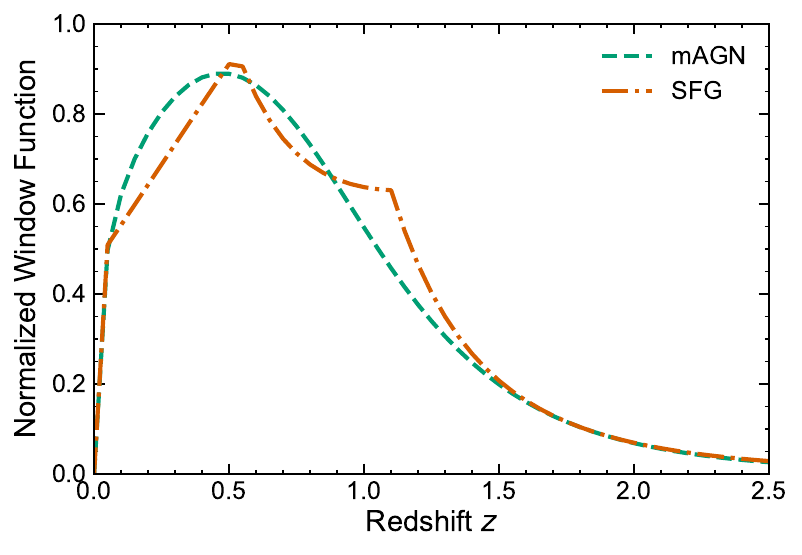}
    \includegraphics[width=0.48\linewidth]{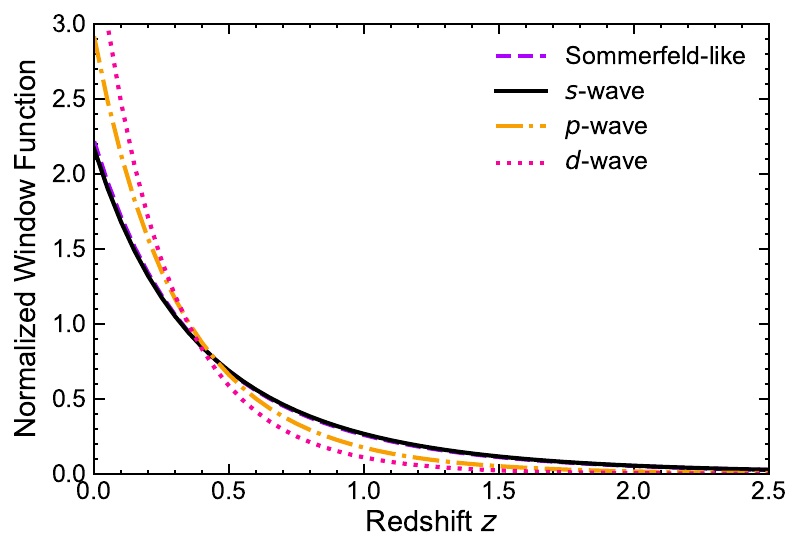}
    \caption{Normalized window functions of the different $\gamma$-ray components as a function of redshift in the $(4.8,\,8.3)\,\mathrm{GeV}$ energy bin. \textit{Left}: The dashed green and dash-dotted brown lines correspond to unresolved mAGNs and SFGs, respectively. \textit{Right}: The Sommerfeld-like (dashed purple), $s$-wave (solid black), $p$-wave (dash-dotted orange), and $d$-wave (dotted pink) annihilation scenarios for a WIMP with $m_{\rm DM}=100\,\mathrm{GeV}$ annihilating through the $b\bar{b}$ channel.}
    \label{fig:WindowFs}
\end{figure*}

The GLF of mAGNs is inferred from the $5\,\mathrm{GHz}$ radio core luminosity function $\phi_{\rm 5\,GHz}$, owing to the limited number of mAGNs detected by current $\gamma$-ray observations \cite{Cholis:2024hmd, Zhou:2024cld}
\begin{equation}\label{GLF_mAGN}
\begin{split}
\phi_{\gamma}(z, L_\gamma) = 
\frac{{\rm d}\ln L_{ \rm 5\,GHz}}
     {{\rm d}\ln L_{\gamma}}
\int & {\rm d}\ln{}L_{ \rm 5\,GHz} \times\\
&
\phi_{ \rm 5\,GHz}(z, L_{ \rm 5\,GHz})
 P(L_{ \rm 5\,GHz}, L_{\gamma}),
\end{split}
\end{equation}
using the empirical correlation
\begin{equation}
    \log_{10}\left(\frac{L_{\gamma}}{\rm erg\,s^{-1}}\right) = 0.78\log_{10}\left(\frac{L_{\rm 5\,GHz}}{10^{40}\,{\rm erg\,s^{-1}}}\right) + 40.78.
\end{equation}
The $P(L_{ \rm 5\,GHz}, L_{\gamma})$ is intrinsic scatter term between the radio and $\gamma$-ray luminosity, and the associated parameters are adopted from Ref.~\cite{2018ApJS..239...33Y}; we do not repeat them here for brevity. For the mAGN spectral index, we take $\Gamma = 2.25$ \cite{Hooper:2016gjy, Blanco:2021icw}, with intrinsic luminosities spanning $10^{40} \le L_{\gamma}/\mathrm{erg\,s^{-1}} \le 10^{50}$. The halo mass-luminosity relation is \cite{Arcari:2022zul}
\begin{equation}
M(L_{\gamma}) = \frac{3.6\times10^{13}\,\mathrm{M_\odot}}{(1+z)^{0.903}} \left(\frac{L_{\gamma}}{10^{48}\,\mathrm{erg\,s^{-1}}}\right)^{0.1032}.
\end{equation}

The normalized window function for unresolved mAGNs in the $(4.8,\,8.3)\,\mathrm{GeV}$ bin is shown as the dashed green line in the left panel of Fig.~\ref{fig:WindowFs}. The corresponding cross-power spectrum with galaxies at $z=0.05$ is displayed as the dashed green curve in the left panel of Fig.~\ref{fig:ThreeD_PS}. Its nearly scale-independent plateau arises because mAGNs are treated as point-like emitters, rendering the one-halo term constant on small scales, in contrast to the scale-dependent galaxy auto-power spectrum.

\subsubsection{Star-forming galaxies}

Similar to mAGNs, the GLF of SFGs is derived from the infrared (IR) luminosity function \cite{Pinetti:2021jjs}
\begin{equation}
\phi_{\gamma}(z, L_\gamma) = \phi_{\rm IR} \, \frac{\mathrm{d}\log_{10}L_{\rm IR}}{\mathrm{d}\log_{10}L_{\gamma}},
\end{equation}
with the $\gamma$-ray-IR luminosity relation \cite{Fermi-LAT:2012nqz}
\begin{equation}
    \log_{10}\left(\frac{L_{\gamma}}{\rm erg\,s^{-1}}\right) = 1.09\log_{10}\left(\frac{L_{\rm IR}}{10^{10}\,{\rm L}_{\odot}}\right) + 39.19.
\end{equation}

The total IR luminosity function is the sum of three subclasses (quiescent spirals, starbursts, and SF-AGN) modeled following Gruppioni et al.~\cite{Gruppioni:2013jna}; the full expression can be found in Ref. \cite{Guo:2026ezl}. For SFGs, we adopt a photon index $\Gamma = 2.7$ \cite{Pinetti:2019ztr} and integrate over $10^{37} \le L_{\gamma}/\mathrm{erg\,s^{-1}} \le \min[10^{42}, L_{\rm th}(z)]$. The halo mass-luminosity relation is \cite{Arcari:2022zul}
\begin{equation}
M(L_{\gamma}) = \frac{10^{12}\,\mathrm{M_\odot}}{(1+z)^{1.61}} \left( \frac{L_{\gamma}}{6.8 \times 10^{39}\,\mathrm{erg\,s^{-1}}} \right)^{0.92}.
\end{equation}
The normalized window function for unresolved SFGs in the $(4.8,\,8.3)\,\mathrm{GeV}$ bin is shown as the dash-dotted brown curve in the left panel of Fig.~\ref{fig:WindowFs}. The multi-peaked structure reflects the distinct redshift evolution of the three SFG subclasses, with quiescent spirals dominate at low redshifts ($z \lesssim 0.5$), SF-AGN peak around $z \simeq 1.1$, and starbursts remain subdominant across all redshifts. The corresponding cross-power spectrum with galaxies at $z=0.05$ is shown as the dash-dotted brown curve in the left panel of Fig.~\ref{fig:ThreeD_PS}.

\subsubsection{Velocity-dependent WIMP DM annihilation}\label{subsec:Velocity-dependent WIMP DM annihilation}
For velocity-dependent WIMP annihilation, we have
\begin{equation}
g_{\rm DM}(\mathbf{x}|z,M) = \rho_{\rm DM}^2(\mathbf{x}|z,M) \, \left (\sigma v_{\rm rel} \right)(\mathbf{x}|z,M),
\end{equation}
where $\left (\sigma v_{\rm rel} \right)$ is the thermally averaged annihilation cross section. To systematically model the velocity dependence, we parameterize the thermally averaged annihilation cross section as
\begin{equation}
\left (\sigma v_{\rm rel} \right) = (\sigma v) \int \mathrm{d}^3\mathbf{v}_1 \, \mathrm{d}^3\mathbf{v}_2 \,
|\mathbf{v}_1 - \mathbf{v}_2|^{n} \, f(\mathbf{v}_1) f({\mathbf{v}_2}),
\end{equation}
where $(\sigma v)$ is a velocity-independent normalization factor to be constrained, and the power-law index $n = -1, 0, 2, 4$ corresponds to Sommerfeld-like, $s$-wave, $p$-wave, and $d$-wave annihilation scenarios, respectively. Assuming an isotropic Maxwell-Boltzmann (MB) velocity distribution $f({\mathbf{v}})$ within virialized halos, the integral evaluates to (see Appendix~\ref{App:WIMP_WF} for the full derivation)
\begin{equation}
\left (\sigma v_{\rm rel} \right)(r|z,M) = (\sigma v) \times
\begin{cases}
\sqrt{\dfrac{2}{\pi}} \dfrac{1}{v_{\rm p}(r|z,M)}, & n = -1, \\
1, & n = 0, \\
3 \, v_{\rm p}^{2}(r|z,M), & n = 2, \\
15 \, v_{\rm p}^{4}(r|z,M), & n = 4,
\end{cases}
\end{equation}
where $v_{\rm p}(r|z,M)$ represents the local most probable speed of DM particles at radius $r$. It is related to the one-dimensional velocity dispersion $\sigma_v$ through $v_{\rm p}^2 = 2\sigma_v^2$. Under the assumption of spherical symmetry and isotropic velocity dispersion, the latter is determined by solving the Jeans equation
\begin{equation}
\sigma_v^2(r|z,M) = \frac{1}{\rho_{\rm DM}(r|z,M)}
\int_{r}^{\infty} \rho_{\rm DM}\, \frac{\mathrm{d}\Phi}{\mathrm{d}s} \, \mathrm{d}s,
\end{equation}
with $\Phi(r|z,M)$ denoting the total gravitational potential of the halo.

\begin{figure}
    \centering
    \includegraphics[width=1.\linewidth]{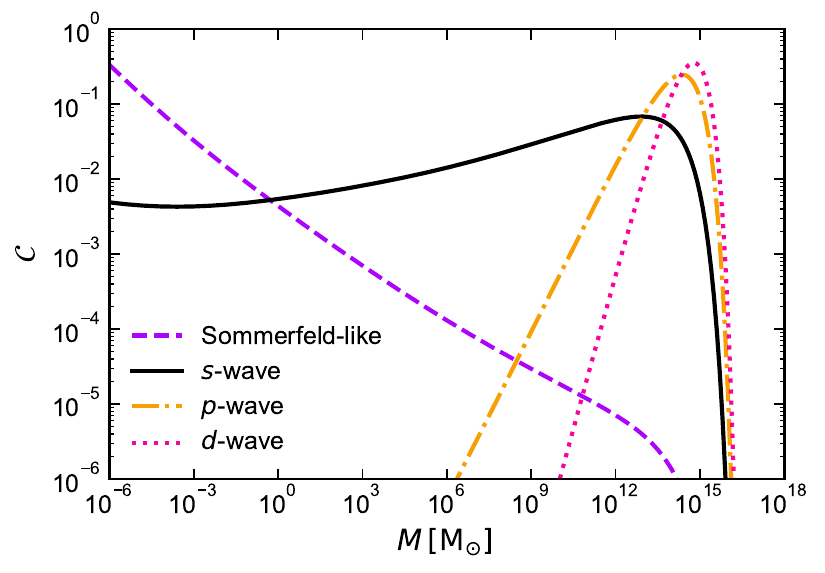}
    \caption{Fractional contribution, $\mathcal{C}(z,M)$, to the velocity-dependent annihilation signal per logarithmic halo mass interval at $z=0.05$ for the Sommerfeld-like (dashed purple), $s$-wave (solid black), $p$-wave (dash-dotted orange), and $d$-wave (dotted pink) annihilation scenarios.}
    \label{fig:Distribution_of_C}
\end{figure}

The theoretical window function governing the prompt DM annihilation contribution to the UGRB is given by
\begin{equation}
W_{\rm DM}(E_{\gamma},\chi) = \frac{\langle g_{\rm DM} \rangle(z)}{8\pi (1+z)^3 m_{\rm DM}^2}
\frac{\mathrm{d}N_{\gamma}^{f}}{\mathrm{d}E_{\gamma}}\big[(1+z)E_{\gamma}\big] \,
e^{-\tau([1+z]E_{\gamma})},
\end{equation}
where $m_{\rm DM}$ is the WIMP mass, $\mathrm{d}N_{\gamma}^{f}/\mathrm{d}E_{\gamma}$ is the prompt $\gamma$-ray spectrum per annihilation event for channel $f$ obtained from {\tt PPPC4} \cite{Cirelli:2010xx}. The averaged quantity is given by
\begin{equation}
\begin{aligned}
    \langle g_{\rm DM} \rangle(z)  =  &\int{\rm d}M\frac{{\rm d}n}{{\rm d}M}\int {\rm d}^{3}\mathbf{x}\,\rho_{\rm DM}^{2}(\mathbf{x}|z, M)\left (\sigma v_{\rm rel} \right)(\mathbf{x}|z,M).
\end{aligned}
\end{equation}

\begin{table*}[t]
\centering
\caption{Best-fit parameter values of the halo boost factor $B(M, z)$. 
These values are adopted from \cite{Moline:2016pbm}.}
\label{tab:BoostF_parameters}
\setlength{\tabcolsep}{0.6em} 
\begin{tabular}{lc c c c c c c} 
\toprule
Parameter & $b_{0}$ & $b_{1}$ & $b_{2}$ & $b_{3}$ & $b_{4}$ & $b_{5}$ \\
\midrule
Value & $-0.186$ & $0.144$ & $-8.8\times10^{-3}$ & $1.13\times 10^{-3}$ & $-3.7\times 10^{-5}$ & $-2\times 10^{-7}$ \\
\bottomrule
\end{tabular}
\end{table*}

To identify which halo mass scales dominate the total annihilation emission across different velocity-dependent regimes, we define the fractional contribution per logarithmic halo mass interval as
\begin{equation}
\begin{aligned}
    \mathcal{C}(z,M) = & \frac{M}{\langle g_{\rm DM} \rangle}\frac{{\rm d}n}{{\rm d}M}\int {\rm d}^{3}\mathbf{x}\,\rho_{\rm DM}^{2}(\mathbf{x}|z, M)\left (\sigma v_{\rm rel} \right)(\mathbf{x}|z,M)
\end{aligned}
\end{equation}
which, by definition, satisfies the normalization condition $\int \mathrm{d}\ln M \, \mathcal{C}(z,M) = 1$.

Figure~\ref{fig:Distribution_of_C} shows $\mathcal{C}(z,M)$ at $z=0.05$ for the four annihilation scenarios. The $s$-wave case (solid black curve) exhibits a broad mass distribution, reflecting the fact that the annihilation rate scales solely with $\rho_{\rm DM}^2$, which receives comparable contributions from low-mass (abundant) and high-mass (dense) halos. In contrast, the $p$-wave (dash-dotted orange) and $d$-wave (dotted pink) scenarios shift the distribution toward more massive halos, where higher velocity dispersions significantly boost the annihilation rate, more than compensating for their lower abundance.

The normalized window functions for each velocity-dependent annihilation scenario, with $m_{\rm DM} = 100\,{\rm GeV}$ via the $b\bar{b}$ channel in the energy bin $(4.8,8.3)\,{\rm GeV}$, are displayed in the right panel of Fig.~\ref{fig:WindowFs}. For the $p$-wave (dash-dotted orange) and $d$-wave (dotted pink) annihilation, the observed $\gamma$-ray emission receives its primary contribution from lower redshifts, where cosmological structure growth yields a higher abundance of massive halos.

The Fourier transform of the field $g_{\rm DM}$, required for the power-spectrum calculation, is given by
\begin{equation}
\begin{aligned}
    \tilde{g}_{\rm DM}(k|z, M)  = & \int {\rm d}^{3}\mathbf{x}\,g_{\rm DM}(\mathbf{x}|z, M)e^{-i\mathbf{k}\cdot \mathbf{x}}\\
    = & \frac{4\pi r_{\rm s}^{6}\rho_{\rm s}^{2}}{k} \int_0^{r_{\rm vir}} \mathrm{d}r \, \frac{\sin(kr)}{r\left(r_{\rm s} + r\right)^{4}} \, \left(\sigma v_{\rm rel}\right)(r|z, M).
\end{aligned}
\end{equation}

\begin{figure}
    \centering
    \includegraphics[width=1.\linewidth]{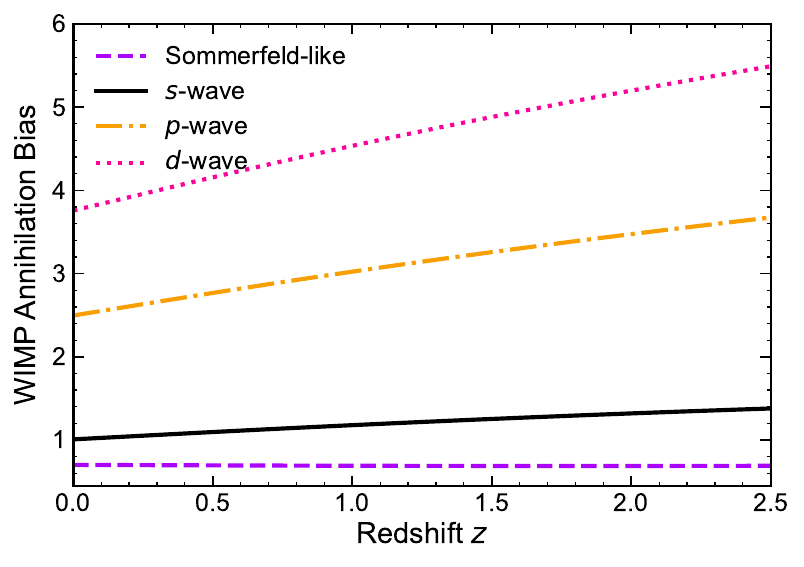}
    \caption{Effective bias, $b_{\rm DM}(z)$, of the WIMP annihilation as a function of redshift for the Sommerfeld-like (dashed purple), $s$-wave (solid black), $p$-wave (dash-dotted orange), and $d$-wave (dotted pink) annihilation scenarios.}
    \label{fig:Bias_WIMP}
\end{figure}

The effective bias of the annihilation field, which characterizes the large-scale clustering amplitude of the DM signal relative to the underlying dark matter distribution, is defined as
\begin{equation}
    b_{\rm DM}(z) = \int {\rm d}M \frac{{\rm d}n}{{\rm d}M}b_{\rm h}(z,M)\int {\rm d}^{3}\mathbf{x}\frac{g_{\rm DM}}{\langle g_{\rm DM}\rangle(z)}.
\end{equation}

Figure~\ref{fig:Bias_WIMP} illustrates the redshift evolution of the effective bias across the four annihilation scenarios. The $p$-wave (dash-dotted orange) and $d$-wave (dotted pink) channels exhibit substantially larger effective biases than the $s$-wave case (solid black). This behavior directly reflects the mass-weighting patterns shown in

Fig.~\ref{fig:Distribution_of_C}. Because halo bias $b_{\rm h}(z,M)$ increases monotonically with halo mass, scenarios whose signals are preferentially sourced by massive halos naturally yield larger effective biases. This effect is particularly pronounced for the $d$-wave case, where the steep velocity dependence ($v_{\rm rel}^4$) strongly favors the high-velocity environments of the most massive halos. In contrast, the $s$-wave scenario yields an effective bias close to unity, indicating that the spatial clustering of the $s$-wave annihilation field closely tracks the underlying linear matter distribution. Conversely, the Sommerfeld-like scenario (dashed purple) produces an effective bias below unity. This anti-biased behavior stems from the signal being dominated by low-mass halos, which are intrinsically less biased tracers of the large-scale structure, resulting in a DM annihilation background that is more smoothly distributed and less clustered than the matter field.

\begin{figure*}[t]
    \centering
    \includegraphics[width=0.48\linewidth]{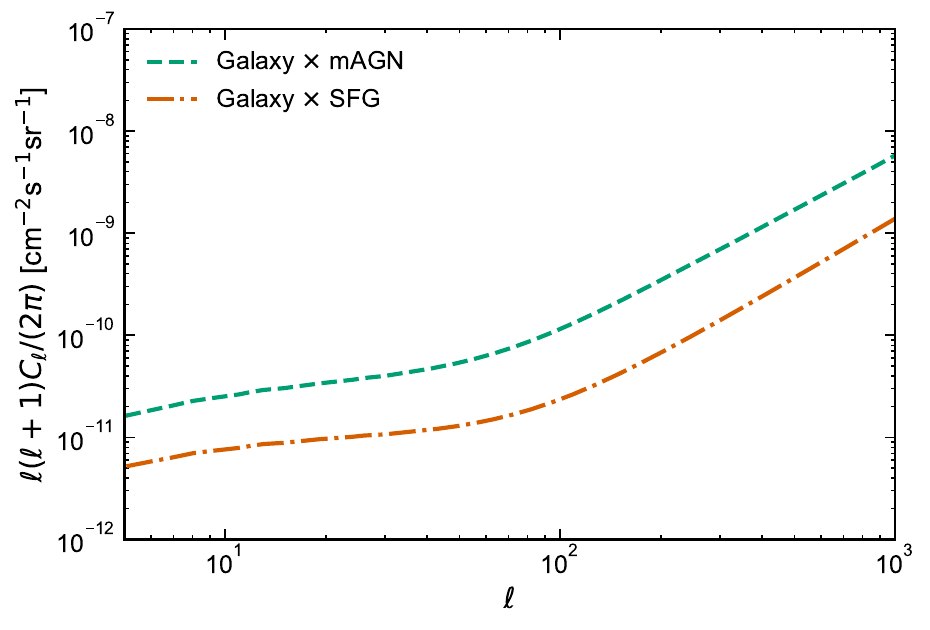}
    \includegraphics[width=0.48\linewidth]{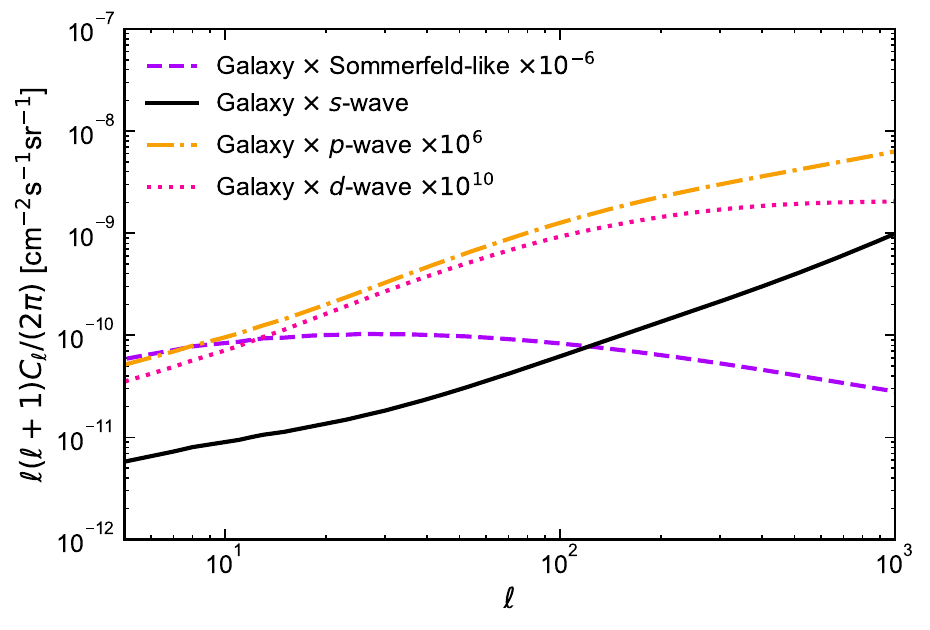}
    \caption{Cross angular power spectra integrated over the redshift range $0 \le z \le 0.1$ in the $(4.8,\,8.3)\,\mathrm{GeV}$ energy bin. \textit{Left}: Contributions from unresolved mAGNs (dashed green) and SFGs (dash-dotted brown). \textit{Right}: Cross-APSs for the Sommerfeld-like (dashed purple), $s$-wave (solid black), $p$-wave (dash-dotted orange), and $d$-wave (dotted pink) WIMP annihilation scenarios, assuming $m_{\rm DM}=100\,\mathrm{GeV}$, the $b\bar{b}$ annihilation channel, and a normalization coefficient $(\sigma v)=3\times10^{-26}\,\mathrm{cm^{3}\,s^{-1}}$. For visual clarity, the amplitudes of the Sommerfeld-like, $p$-wave, and $d$-wave curves have been rescaled.}
    \label{fig:APS}
\end{figure*}

The resulting cross-power spectra for the Sommerfeld-like (dashed purple), $s$-wave (solid black), $p$-wave (dash-dotted orange), and $d$-wave (dotted pink) annihilation scenarios (cross-correlated with the galaxy catalog at $z=0.05$) are presented in the right panel of Fig.~\ref{fig:ThreeD_PS}. The four velocity-dependent models exhibit distinct clustering behaviors. Velocity dependence modifies the annihilation field through two complementary mechanisms. First, it changes the relative contribution of halos of different masses through the velocity-dependent annihilation kernel, thereby altering the effective halo bias and the overlap between the annihilation field and the galaxy population. Second, it reshapes the radial emissivity profile within individual halos, leading to distinct Fourier-space behaviors of the one-halo term.

For the Sommerfeld-like scenario, the annihilation rate is enhanced in low-velocity environments and is consequently dominated by low-mass halos. Since halos below the HOD mass threshold contain essentially few or no galaxies, the overlap between the annihilation field and the galaxy distribution within the same halo is strongly suppressed, rendering the one-halo contribution nearly negligible. Furthermore, as discussed above, the effective bias for the Sommerfeld-like case is correspondingly reduced, yielding the weakest two-halo term among all velocity dependences.

Conversely, the $p$-wave and particularly the $d$-wave scenarios preferentially weight massive halos characterized by large velocity dispersions. These massive systems naturally host abundant galaxies, especially satellite populations, fostering a strong intra-halo overlap between the galaxy and DM annihilation fields, which substantially bolsters the one-halo term. Simultaneously, because massive halos are highly biased tracers of the underlying matter field, the effective bias increases systematically from $s$-wave to $p$-wave and $d$-wave scenarios, naturally producing progressively larger two-halo spectrum amplitudes.

Interestingly, the scale dependence of the one-halo term originates from a distinct physical mechanism. The Fourier transform of the annihilation field is determined by the radial emissivity profile, which can be treated as a velocity-weighted squared dark matter profile. In NFW halos, the velocity dispersion vanishes at the center and peaks near the scale radius $r_{\rm s}$. Consequently, higher powers of velocity weighting ($n=2, 4$) suppress the central emission peak relative to the outer regions, effectively broadening the radial profile of the annihilation field in real space. According to the properties of the Fourier transform, a spatially more extended real-space profile corresponds to a steeper profile in Fourier space, causing $\tilde{g}_{\rm DM}(k)$ to decrease more rapidly with increasing $k$. This naturally explains why the one-halo contribution of the $d$-wave model exhibits the steepest decline towards small scales, followed by the $p$-wave and finally the $s$-wave scenario.

We note that all calculations presented above account for the contribution from subhalos within the primary halo by replacing $\rho_{\rm DM}^2$ with $[1 + B(z,M)]\,\rho_{\rm DM}^2$, where $B(z,M)$ is the subhalo boost factor. We adopt the parameterization from Ref.~\cite{Moline:2016pbm}
\begin{equation}
\log_{10} B(z,M) = \sum_{k=0}^{5} b_k \left[ \log_{10}\left(\frac{M}{\mathrm{M_\odot}}\right) \right]^k,
\end{equation}
with the best-fit coefficients $b_k$ listed in Table~\ref{tab:BoostF_parameters}.

\section{Sensitivity Forecast and Discussion}\label{Sec:ResultDiscussion}
Using the window functions and three-dimensional power spectra derived in the preceding section, we calculate the cross APS between the components contributing to the UGRB and the 2MRS galaxy catalog using Eq.~(\ref{eq:APS}).

Figure~\ref{fig:APS} displays the predicted cross-APS integrated over the redshift interval $0 \le z \le 0.1$ within the energy bin $(4.8,8.3)\,\mathrm{GeV}$. The left panel shows the dominant astrophysical contributions to the $\gamma$-ray emission from unresolved mAGNs (dashed green) and SFGs (dash-dotted brown). The right panel illustrates the cross-correlation signals for the four velocity-dependent WIMP annihilation scenarios, assuming a benchmark WIMP mass of $m_{\rm DM} = 100\,\mathrm{GeV}$ annihilating via the $b\bar{b}$ channel with a normalization factor of $(\sigma v) = 3\times 10^{-26}\,{\rm cm^3\,s^{-1}}$. For visual clarity, the amplitudes of the Sommerfeld-like, $p-$, and $d$-wave annihilation scenarios in the right panel have been rescaled.

To estimate the statistical uncertainty of the predicted cross-APS, we adopt a Gaussian covariance model
\begin{equation}
    \left(\Delta C_{\ell}^{\gamma\mathrm{g}}\right)^{2} = \frac{f_{\mathrm{sky}}^{-1}}{2\ell + 1} \left[ \left(C_{\ell}^{\gamma\mathrm{g}}\right)^{2} + \left(C_{\ell}^{\gamma\gamma} + \frac{C_{\mathrm{N}}^{\gamma}}{\left(B_{\ell}^{\gamma}\right)^{2}}\right) \left(C_{\ell}^{\mathrm{gg}} + C_{\mathrm{N}}^{\mathrm{g}}\right) \right],
\end{equation}
where $f_{\mathrm{sky}} = \min\{f_{\mathrm{2MRS}}, f_{\mathrm{Fermi}}\}$ represents the effective overlapping sky fraction. Here, $C_{\ell}^{\gamma\gamma}$ and $C_{\ell}^{\mathrm{gg}}$ denote the auto-correlation APS of the $\gamma$-ray sky and the galaxy distribution, respectively. The term $C_{\mathrm{N}}^{\gamma}$ accounts for $\gamma$ photon Poisson noise (summarized in Table~\ref{tab:FermiLAT_specifications}), and $C_{\mathrm{N}}^{\mathrm{g}} = 4\pi f_{\mathrm{2MRS}} / N_{\mathrm{g}}$ quantifies the galaxy shot noise.

The finite angular resolution of the \textit{Fermi}-LAT instrument suppresses anisotropy power on small angular scales. We account for this attenuation via the energy-dependent beam window function
\begin{equation}
    B_{\ell}^{\gamma}(E_\gamma) = \exp \left[ -\frac{\theta_b^2(E_\gamma, \ell) \ell^2}{2} \right],
\end{equation}
where the effective beam size $\theta_b(E_\gamma, \ell)$ is parameterized as
\begin{equation}
    \theta_b(E_\gamma, \ell) = \theta_{\mathrm{cont}}^{68\%}(E_\gamma) \left[ 1 + 0.25 \, \theta_{\mathrm{cont}}^{68\%}(E_\gamma) \ell \right]^{-1}.
\end{equation}
Following Refs.~\cite{Pinetti:2019ztr,Zhou:2024cld}, the $68\%$ containment angle is modeled as
\begin{equation}
    \theta_{\mathrm{cont}}^{68\%}(E_\gamma) = \theta_{\mathrm{ref}}^{68\%} \left( \frac{E_\gamma}{E_{\mathrm{ref}}} \right)^{-0.95} + 0.05\, {\rm deg} ,
\end{equation}
with reference parameters $E_{\mathrm{ref}} = 0.5\,\mathrm{GeV}$ and $\theta_{\mathrm{ref}}^{68\%} = 1.20\,{\rm deg}$. Since the angular resolution of modern galaxy surveys is substantially superior to that of $\gamma$-ray telescopes, no beam correction is required for the galaxy density field.

\begin{figure*}
    \centering
    \includegraphics[width=1.\linewidth]{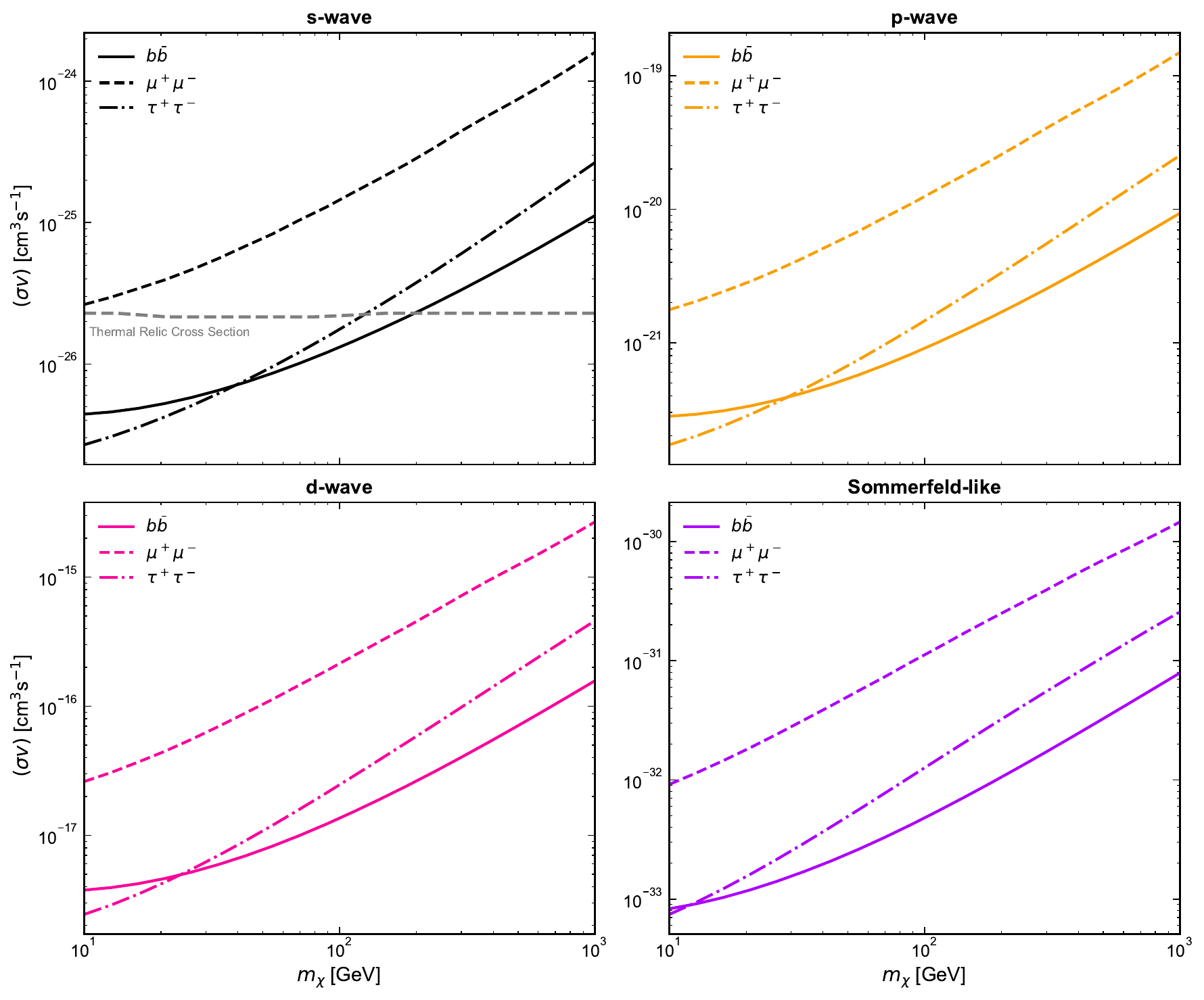}
    \caption{Projected $2\sigma$ ($95\%$ confidence level) upper limits on the velocity-independent annihilation coefficient $(\sigma v)$ as a function of the WIMP mass. The upper-left, upper-right, lower-left, and lower-right panels correspond to the $s$-wave, $p$-wave, $d$-wave, and Sommerfeld-like annihilation scenarios, respectively. In each panel, the solid, dashed, and dash-dotted curves denote the projected limits for the $b\bar{b}$, $\mu^{+}\mu^{-}$, and $\tau^{+}\tau^{-}$ annihilation channels, respectively. In the $s$-wave panel, the canonical thermal relic cross section is shown by the gray dashed line for comparison \cite{Steigman:2012nb}.}
    \label{fig:UpperLimits}
\end{figure*}

To evaluate the projected sensitivity to the WIMP annihilation cross section, we perform a statistical forecast. The null hypothesis assumes that the observed cross-correlation signal is entirely sourced by unresolved astrophysical backgrounds (mAGNs and SFGs), whereas the alternative hypothesis incorporates the additional WIMP-induced $\gamma$-ray component. Following Refs.~\cite{Pinetti:2019ztr, Zhou:2024cld}, the test statistic $\Delta\chi^2$ for a given WIMP mass is quantified by 
\begin{equation}
\Delta\chi^2 = \sum_{\ell,e} \left( \frac{C_{\ell,e}^{\gamma\mathrm{g}}}{\Delta C_{\ell,e}^{\gamma\mathrm{g}}} \right)^2_{\star+\mathrm{WIMP}} - \sum_{\ell,e} \left( \frac{C_{\ell,e}^{\gamma\mathrm{g}}}{\Delta C_{\ell,e}^{\gamma\mathrm{g}}} \right)^2_{\star},
\end{equation}
where the summation runs over all angular multipoles $\ell$ and $\gamma$-ray energy bins $e$.

For each velocity-dependent scenario, the velocity-independent annihilation coefficient $(\sigma v)$ is the sole additional free parameter for a fixed WIMP mass. As a result, $\Delta\chi^2$ asymptotically follows a $\chi^2$ distribution with one degree of freedom. We adopt $\Delta\chi^2 = 4$ to derive the projected $95\%$ confidence level ($2\sigma$) exclusion limits.

The resulting projected constraints on $(\sigma v)$ across different velocity-dependent scenarios are presented in Fig.~\ref{fig:UpperLimits}. For completeness, in addition to the benchmark hadronic channel ($b\bar{b}$), we also report upper limits for the leptonic annihilation channels ($\mu^+\mu^-$ and $\tau^+\tau^-$). Previous studies have extensively constrained velocity-dependent DM annihilation using $\gamma$-ray observations of local dwarf spheroidal galaxies (dSphs; e.g., Ref.~\cite{Boddy:2019qak}). However, dSphs reside in low-mass subhalos ($M \sim 10^6\text{--}10^8\,\mathrm{M_\odot}$) with low velocity dispersions ($\sigma_{v} \sim 10\,\mathrm{km/s}$). As demonstrated in our halo mass decomposition analysis, while such low-velocity environments favor Sommerfeld-enhanced annihilation ($n=-1$), they severely suppress $p$-wave ($n=2$) and $d$-wave ($n=4$) channels, rendering dSphs suboptimal targets for testing velocity-dependent DM annihilation models. In contrast, cross-correlations between the UGRB and large-scale galaxy catalogs integrate signals across the entire halo mass range, fully exploiting both high-dispersion environments in massive halos and low-dispersion ones in low-mass halos. Consequently, our large-scale structure cross-correlation approach offers significantly enhanced sensitivity for velocity-dependent annihilation channels compared to traditional dwarf galaxy searches.

The adopted minimum halo mass, $M_{\rm min}$, constitutes one of the dominant theoretical uncertainties in predictions of velocity-dependent dark matter annihilation. While the predicted signals for the $s$-, $p$-, and $d$-wave scenarios are largely insensitive to the choice of $M_{\rm min}$, the Sommerfeld-like annihilation signal exhibits a pronounced dependence on the adopted cutoff mass. Specifically, increasing $M_{\rm min}$ from $10^{-6}\,{\rm M_{\odot}}$ to $1\,{\rm M_{\odot}}$ suppresses the predicted signal by approximately two orders of magnitude. Since the minimum halo mass depends on the microphysical properties of dark matter, in particular the kinetic decoupling temperature, the associated uncertainty should therefore be regarded as an intrinsic theoretical uncertainty of Sommerfeld-enhanced annihilation models.

{A related uncertainty arises from the extrapolation of the concentration--mass relation. The $c_{200}$--$M$ relation adopted here is calibrated over $10^{-2}\lesssim M/\mathrm{M_\odot}\lesssim10^{16}$, whereas our halo-model integration extends over $10^{-6}\le M/\mathrm{M_\odot}\le10^{18}$, so that the relation is extrapolated over four decades at low masses and two decades at high masses. As shown by the fractional contribution $\mathcal{C}(z,M)$ in Fig.~\ref{fig:Distribution_of_C}, the impact of this extrapolation is $n$-dependent. For the $p$- and $d$-wave scenarios, the signal is dominated by halos of mass around $10^{14}\,\mathrm{M_\odot}$, well within the calibrated range, so the extrapolation has a negligible effect. For the $s$-wave scenario, the signal receives its dominant contribution from halos of mass $M\sim10^{12}\,\mathrm{M_\odot}$, with only a small contribution from lower-mass halos, so the low-mass extrapolation introduces only a minor uncertainty. For the Sommerfeld-like scenario, by contrast, the signal is increasingly dominated by lower-mass halos, so the low-mass extrapolation of the $c_{200}$--$M$ relation has a larger impact. This uncertainty is, however, mitigated by the saturation of the Sommerfeld enhancement for a finite mediator mass, as discussed below.}

{Here, we note that the Sommerfeld-like scenario considered in this work corresponds to the Coulomb limit of Sommerfeld enhancement, in which the mediator is assumed to be massless and the enhancement factor takes the form of $S(v_{\rm rel}) = 2\pi\alpha_\chi/v_{\rm rel}$ in the strongly enhanced regime
$2\pi\alpha_\chi \gg v_{\rm rel}$. Following Ref. \cite{Boddy:2019qak}, we adopt $\alpha_\chi = 1/(2\pi)$, which yields $\sigma v_{\rm rel} = (\sigma v)\, v_{\rm rel}^{-1}$ and allows a direct comparison with existing results in the literature. This parameterization should therefore be regarded as an idealized massless-mediator limit.

For a mediator of finite mass $m_\phi$, the interaction is governed by a Yukawa potential and the Sommerfeld enhancement saturates at $v_{\rm rel} \sim m_\phi/m_\chi$. In this regime, the enhancement of low-mass halos, characterized by low velocities, is suppressed relative to the idealized case considered here. Interestingly, this saturation also mitigates the uncertainty from the low-mass extrapolation of the concentration--mass relation discussed above: since the contribution of the lowest-mass halos is suppressed by saturation, the sensitivity of the Sommerfeld-enhancement signal to the extrapolated concentration values in the range $10^{-6}\lesssim M/(\mathrm{M_\odot})\lesssim10^{-2}$ is reduced. A fully self-consistent treatment of the Sommerfeld enhancement with a finite mediator mass requires solving the Schr\"odinger equation for a Yukawa potential and is inherently model-dependent, depending on $m_\phi$, $\alpha_\chi$, and $m_\chi$, and potentially involving resonant enhancement. We leave such an analysis to future work.
}

Another important source of theoretical uncertainty arises from the treatment of substructure. The boost factor adopted in this work was originally calibrated for velocity-independent ($s$-wave) annihilation and was directly applied to all velocity-dependent scenarios. This approximation implicitly assumes that the subhalo boost is independent of the annihilation velocity dependence. However, subhalos generally possess substantially lower internal velocity dispersions than their host halos. Consequently, their contribution is expected to be suppressed for $p$- and $d$-wave annihilation, while being further enhanced for Sommerfeld-like scenarios \cite{Piccirillo:2022qet, Lacroix:2022cjm}. A fully self-consistent treatment therefore requires modeling the internal phase-space distribution of subhalos and recalibrating the boost factor separately for each annihilation mechanism. Such an extension is beyond the scope of the present work but represents an important direction for future studies.

{Finally, we note that the four annihilation scenarios considered above represent idealized single-partial-wave limits. In realistic models, however, multiple partial waves can contribute, with the $s$-wave component generally suppressed, but not necessarily vanishing, in velocity-suppressed scenarios. To assess the impact of such an admixture, we consider the representative case of $s$--$p$ mixing, parameterized as $\sigma v_{\rm rel} = (\sigma v) \left[\epsilon + v_{\rm rel}^{2}\right]$, where $\epsilon$ is a dimensionless mixing parameter controlling the relative normalization of the $s$- and $p$-wave components. We find that the admixture leads to a mass-dependent transition from $s$-wave dominance in low-mass halos to $p$-wave dominance in high-mass halos. Consequently, the effective bias and clustering properties of the annihilation signal evolve continuously between the pure $s$- and $p$-wave scenarios and cannot, in general, be described by a single power-law index $n$. A detailed analysis of these partial-wave admixture cases is presented in Appendix~\ref{app:admixture_cases}.}

More generally, the present work adopts the generic velocity expansion of the annihilation cross section as a phenomenological description of velocity-dependent dark matter annihilation. The halo-model framework developed here can therefore be readily generalized to specific particle physics scenarios that predict characteristic forms of the velocity dependence, including Sommerfeld-enhanced or resonance-mediated annihilation models. Future cosmological measurements combining spectral and clustering information from large-scale structure tracers are expected to provide increasingly powerful probes of velocity-dependent dark matter interactions.


\section{Conclusion}\label{Sec:Conclusion}

In this work, we have investigated the cosmological signatures of velocity-dependent WIMP dark matter annihilation through the cross-correlation between the unresolved $\gamma$-ray background and large-scale structure. Unlike the conventional velocity-independent ($s$-wave) scenario, velocity-dependent annihilation couples the annihilation rate to the internal kinematics of dark matter halos, implying that different annihilation mechanisms can produce distinct spatial clustering and redshift distributions of the cosmological $\gamma$-ray signal. This motivates the UGRB--LSS cross-correlation as a promising observational probe of the velocity dependence of dark matter annihilation.

To this end, we have developed a generalized halo-model framework for velocity-dependent dark matter annihilation by parameterizing the velocity-weighted annihilation cross section as $\sigma v_{\rm rel}\propto v_{\rm rel}^{n}$. This unified formalism allows us to investigate four representative scenarios, corresponding to Sommerfeld-like ($n=-1$), $s$-wave ($n=0$), $p$-wave ($n=2$), and $d$-wave ($n=4$) annihilation. Within this framework, we computed the UGRB--galaxy catalog cross angular power spectrum and performed sensitivity forecasts for the \textit{Fermi}-LAT and 2MRS cross-correlation, deriving the projected constraints on the velocity-independent annihilation coefficient for different annihilation scenarios.

Our results show that velocity-dependent annihilation modifies not only the overall annihilation rate but also the clustering properties of the cosmological signal. Different velocity dependencies preferentially weight halo populations of different masses, leading to characteristic effective biases and radial emissivity profiles that jointly modulate the UGRB–LSS cross-correlation. In particular, Sommerfeld-like annihilation is dominated by low-mass, weakly biased halos and therefore exhibits substantially weaker clustering. In contrast, $p$-wave and especially $d$-wave annihilation preferentially trace massive halos with large velocity dispersions, substantially enhancing clustering amplitudes. By naturally combining contributions from halos spanning a broad range of masses and velocity dispersions, the cosmological cross-correlation approach provides significantly improved sensitivity to velocity-dependent annihilation scenarios compared to conventional searches based on nearby dwarf spheroidal galaxies.

The framework developed in this work provides a general and flexible approach for incorporating velocity-dependent dark matter annihilation into cosmological halo-model calculations. Although our analysis adopts a phenomenological velocity expansion, the formalism can be readily extended to specific particle physics models predicting more complex forms of velocity dependence. Looking forward, forthcoming large-scale structure surveys together with next-generation $\gamma$-ray observations are expected to substantially improve the sensitivity of cosmological cross-correlation measurements, opening a promising avenue for probing the particle nature of dark matter in the large-scale structure.

\section{Acknowledgments}
The financial assistance of the South African Radio Astronomy Observatory (SARAO) towards this research is hereby acknowledged (\url{www.sarao.ac.za}).

\clearpage

\appendix
\onecolumngrid

\section{Window Function for Velocity-dependent WIMP Annihilation}\label{App:WIMP_WF}
The observed differential $\gamma$-ray intensity along a given line of sight $\mathbf{n}$ can be written in the general form \cite{Fornengo:2013rga}
\begin{equation}\label{eq:gamma_intensity}
    I_{\gamma}(\mathbf{n})=
    \int {\rm d}\chi\,
    \frac{g(\mathbf{n},\chi)}
         {\langle g(\mathbf{n},\chi)\rangle}
    W(\chi),
\end{equation}
where $\chi$ is the comoving distance, $g(\mathbf{n},\chi)$ denotes the source field, $W(\chi)$ is the corresponding window function encoding the redshift-dependent physical contributions to the observed signal, and $\langle\cdots\rangle$ denotes the ensemble average.

The differential $\gamma$-ray intensity observed at energy $E_{\gamma}$ is obtained by integrating the cosmological volume emissivity along the line of sight \cite{Peacock:1999ye, Ando:2005xg}
\begin{equation}\label{eq:app_EI}
I_{\gamma}(\mathbf{n},E_{\gamma})
=
\frac{c}{4\pi E_{\gamma}}
\int
{\rm d}z\,
\frac{
P_{\gamma}\!\left([1+z]E_{\gamma},
\mathbf{n},z\right)}
{H(z)(1+z)^4}
e^{-\tau([1+z]E_{\gamma})},
\end{equation}
where $H(z)$ is the Hubble parameter and $\tau$ is the optical depth due to extragalactic background light absorption.

For WIMP annihilation, the volume emissivity is given by
\begin{equation}\label{eq:app_emissivity}
P_{\gamma}(E_{\gamma})
=
E_{\gamma}
\frac{{\rm d}N_{\gamma}^{f}}{{\rm d}E_{\gamma}}
\,
\frac{\left(\sigma v_{\rm rel}\right)}{2}
\left(
\frac{\rho_{\rm DM}}
     {m_{\rm DM}}
\right)^2,
\end{equation}
where ${\rm d}N_{\gamma}^{f}/{\rm d}E_{\gamma}$ is the prompt $\gamma$-ray spectrum per annihilation event for channel $f$. The factor of $1/2$ accounts for the annihilation of identical dark matter particle pairs. For velocity-dependent annihilation, the thermally averaged annihilation cross section is parameterized as
\begin{equation}
\left(\sigma v_{\rm rel}\right)=(\sigma v)\int {\rm d}^3\mathbf{v}_1 {\rm d}^3\mathbf{v}_2\, |\mathbf{v}_1-\mathbf{v}_2|^n f(\mathbf{v}_1) f(\mathbf{v}_2),
\end{equation}
where $(\sigma v)$ is velocity-independent normalization coefficient.

Assuming that dark matter particles follow an isotropic Maxwell--Boltzmann velocity distribution within virialized halos
\begin{equation}
f(\mathbf{v}|r)\, {\rm d}^3\mathbf{v} = \left( \frac{1} {\pi v_{\rm p}^2(r)}\right)^{3/2} \exp \left[ - \frac{|\mathbf{v}|^2} {v_{\rm p}^2(r)} \right] {\rm d}^3\mathbf{v},
\end{equation}
where $v_{\rm p}(r|z,M)$ is the local most probable speed, we introduce the relative and center-of-mass velocities
\begin{equation}
\mathbf{v}_{\rm rel}
=
\mathbf{v}_1-\mathbf{v}_2,
\qquad
\mathbf{V}
=
\frac{\mathbf{v}_1+\mathbf{v}_2}{2},
\end{equation}
whose Jacobian is unity. The integration then becomes
\begin{equation}
\begin{aligned}
&
\int
{\rm d}^3\mathbf{v}_1
{\rm d}^3\mathbf{v}_2\,
|\mathbf{v}_1-\mathbf{v}_2|^n
f(\mathbf{v}_1)
f(\mathbf{v}_2)
\\
=&
\left(
\frac{1}{\pi v_{\rm p}^2}
\right)^3
\int
{\rm d}^3\mathbf{v}_{\rm rel}
{\rm d}^3\mathbf{V}
\,
|\mathbf{v}_{\rm rel}|^n
\exp
\left[
-
\frac{
2V^2+v_{\rm rel}^2/2}
{v_{\rm p}^2}
\right]
\\
=&
\left(
\frac{1}
{2\pi v_{\rm p}^2}
\right)^{3/2}
4\pi
\int_0^{\infty}
{\rm d}v\,
v^{n+2}
\exp
\left(
-
\frac{v^2}
{2v_{\rm p}^2}
\right)
\\
=&
\frac{
2^{(n+2)/2}}
{\sqrt{\pi}}
\Gamma
\left(
\frac{n+3}{2}
\right)
v_{\rm p}^n
\\
=&
\begin{cases}
\displaystyle
\sqrt{\dfrac{2}{\pi}}
\dfrac{1}{v_{\rm p}},
&
n=-1,
\\[0.3cm]
1,
&
n=0,
\\[0.2cm]
3v_{\rm p}^2,
&
n=2,
\\[0.2cm]
15v_{\rm p}^4,
&
n=4.
\end{cases}
\end{aligned}
\end{equation}
Strictly speaking, the upper limit of the velocity integral should be the local escape velocity rather than infinity. Throughout this work, we adopt the commonly used approximation of extending the integration to infinity, which considerably simplifies the calculation while introducing only a negligible correction for virialized halos.
Since the most probable speed varies with position inside a halo, the thermally averaged annihilation cross section is likewise position dependent, which we denote by $\left(\sigma v_{\rm rel}\right)(\mathbf{x}|z,M)$. Using the relation ${\rm d}\chi=c\,{\rm d}z/H(z)$ and substituting Eq.~(\ref{eq:app_emissivity}) into Eq.~(\ref{eq:app_EI}), we obtain
\begin{equation}
I_{\gamma}(E_{\gamma})
=
\int
\frac{{\rm d}\chi}{8\pi}
\,
\frac{
\left(\sigma v_{\rm rel}\right)
\rho_{\rm DM}^2}
{(1+z)^3m_{\rm DM}^2}
\,
\frac{{\rm d}N_{\gamma}^{f}}
{{\rm d}E_{\gamma}}
\!\left[(1+z)E_{\gamma}\right]
e^{-\tau([1+z]E_{\gamma})}.
\end{equation}

\begin{figure}
    \centering
    \includegraphics[width=0.5\linewidth]{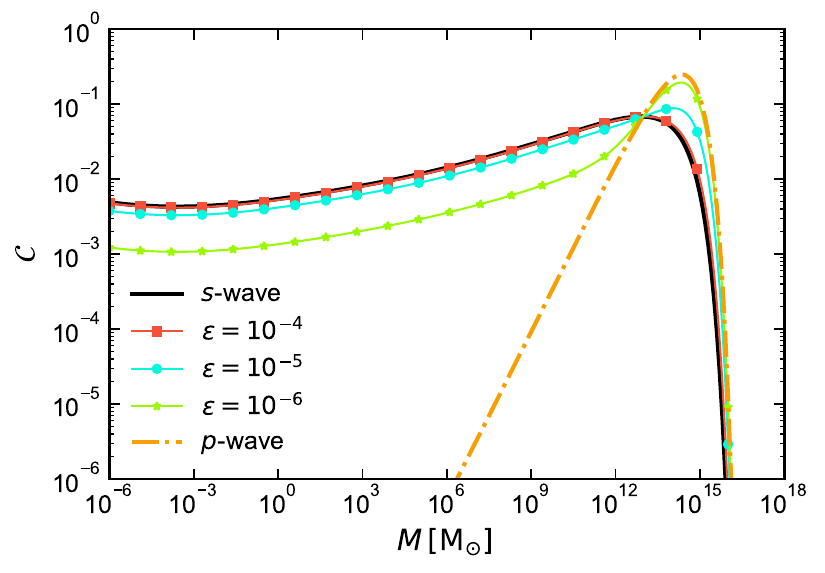}
    \caption{Fractional contribution, $\mathcal{C}(z,M)$, to the velocity-dependent annihilation signal per logarithmic halo mass interval at $z=0.05$. The $s$-$p$ mixing cases with $\epsilon=10^{-4}$, $10^{-5}$, and $10^{-6}$ are shown by the red squares, cyan circles, and lime-green stars, respectively. The $s$-wave and $p$-wave cases are shown by the solid black and dash-dotted orange curves, respectively, for comparison.}
    \label{fig:Distribution_of_C_spwave}
\end{figure}

Defining the spatial intensity field as
\begin{equation}
g_{\rm DM}(\mathbf{x}|z,M)
=
\rho_{\rm DM}^2(\mathbf{x}|z,M)
\left(\sigma v_{\rm rel}\right)(\mathbf{x}|z,M),
\end{equation}
and comparing the above expression with Eq.~(\ref{eq:gamma_intensity}), we identify the corresponding window function for velocity-dependent WIMP annihilation as
\begin{equation}
W_{\rm DM}(E_{\gamma},\chi)
=
\frac{\langle g_{\rm DM}\rangle(z)}
{8\pi(1+z)^3m_{\rm DM}^2}
\frac{{\rm d}N_{\gamma}^{f}}
{{\rm d}E_{\gamma}}
\!\left[(1+z)E_{\gamma}\right]
e^{-\tau([1+z]E_{\gamma})}.
\end{equation}

{
\section{Admixture cases}\label{app:admixture_cases}
We generalize the velocity-dependent annihilation cross section to include 
both $s$- and $p$-wave contributions
\begin{equation}
\sigma v_{\rm rel} = (\sigma v) \left[\epsilon + v_{\rm rel}^{2}\right]
\label{eq:admix}
\end{equation}
where $(\sigma v)$ is a velocity-independent normalization factor and  $\epsilon$ is a dimensionless mixing parameter controlling the relative normalization of the $s$- and $p$-wave components.
Consequently, the thermally averaged annihilation cross section can be expressed as
\begin{equation}
    \left( \sigma v_{\rm rel} \right) = (\sigma v) \left[\epsilon + 3v_{\rm p}^{2}(r|z, M)\right].
\end{equation}
For $\epsilon=0$, the annihilation is purely $p$-wave. In the opposite limit, $\epsilon\gg3v_{\rm p}^{2}(r|z,M)$, the $s$-wave contribution dominates locally and the result approaches the pure $s$-wave limit. For intermediate values, the relative importance of the two partial waves depends on the local velocity dispersion through the ratio $\epsilon/[3v_{\rm p}^{2}(r|z,M)]$. Consequently, a fixed partial-wave admixture can lead to different effective $s$- and $p$-wave contributions across radius, halo mass, and redshift.

We consider three representative cases with $\epsilon=10^{-4}$, $10^{-5}$, and $10^{-6}$. Figure~\ref{fig:Distribution_of_C_spwave} shows $\mathcal{C}(z,M)$ at $z=0.05$, which quantifies the fractional contribution to the total annihilation emission per logarithmic halo-mass interval. The cases with $\epsilon=10^{-4}$, $10^{-5}$, and $10^{-6}$ are shown by the red squares, cyan circles, and lime-green stars, respectively, while the pure $s$-wave and $p$-wave cases are shown by the solid black and dash-dotted orange curves for comparison.

The resulting halo-mass dependence reflects the competition between the $s$-wave term, $\epsilon$, and the $p$-wave term, $3v_{\rm p}^{2}(r|z,M)$. For $\epsilon=10^{-4}$, the 
$s$-wave term dominates for the majority of halos, and the $p$-wave contribution becomes important only in the most massive halos, where $3v_{\rm p}^{2}\gg\epsilon$. Since such halos are relatively rare, the resulting distribution closely follows that of the $s$-wave case. As $\epsilon$ decreases, the $p$-wave term becomes increasingly important across a broader range of halo masses. For $\epsilon=10^{-6}$, the $p$-wave contribution becomes significant in massive halos, while the $s$-wave term remains dominant in lower-mass halos, resulting in a pronounced intermediate behavior between the pure $s$- and $p$-wave limits. The $\epsilon=10^{-5}$ case lies between these two limits, exhibiting a
$p$-wave contribution that is more extended than in the
$\epsilon=10^{-4}$ case but less pronounced than in the
$\epsilon=10^{-6}$ case.

For completeness, the resulting three-dimensional cross-power spectra between the WIMP annihilation field and the galaxy distribution at $z=0.05$ are shown in the left panel of Fig.~\ref{fig:CPS_bias_spwave}. The cross-power spectra for the $s$--$p$ admixture cases reflect the combined contributions of the $s$- and $p$-wave components, whose effects on the halo weighting and clustering properties were discussed in Section~\ref{subsec:Velocity-dependent WIMP DM annihilation}. The corresponding redshift evolution of the effective bias is shown in the right panel. For $\epsilon=10^{-4}$, the effective bias remains close to the pure $s$-wave case, with a small deviation toward the $p$-wave behavior at low redshifts. As $\epsilon$ decreases to $10^{-5}$ and $10^{-6}$, the effective bias increasingly approaches the $p$-wave result toward low redshifts. At higher redshifts, the effective bias instead tends toward the $s$-wave behavior. This redshift dependence reflects the hierarchical growth of structure: massive halos become increasingly abundant toward lower redshifts, and their larger velocity dispersions enhance the $p$-wave term, driving the effective bias toward the pure $p$-wave behavior. At higher redshifts, the signal is dominated by lower-mass halos with smaller velocity dispersions, so the $s$-wave term remains relatively more important and the effective bias approaches the pure $s$-wave result.

Finally, we compute the corresponding angular cross-power spectra for the three representative admixture cases, $\epsilon=10^{-4}$, $10^{-5}$, and $10^{-6}$, shown in the left panel of Fig.~\ref{fig:APS_UpperLimits_spwave}. The spectra are obtained by integrating over $0\leq z\leq0.1$ and are shown for the $(4.8,8.3)\,\mathrm{GeV}$ energy bin. We adopt a benchmark WIMP mass of $m_{\rm DM}=100\,\mathrm{GeV}$, annihilation into the $b\bar b$ channel, and a normalization of $(\sigma v)=3\times10^{-26}\,{\rm cm^3s^{-1}}$. For comparison, the pure $s$- and $p$-wave cases are also shown by the solid black and dash-dotted orange curves, respectively.

In the right panel, we present the projected $95\%$ confidence-level upper limits on $(\sigma v)$ for three fixed values of $\epsilon=10^{-4}$, $10^{-5}$, and $10^{-6}$, together with the corresponding constraints for the pure $s$- and $p$-wave scenarios. Here $(\sigma v)$ is treated as the sole free parameter for each fixed $\epsilon$. A joint two-parameter analysis of $\epsilon$ and $(\sigma v)$ in two dimensions is left for future work.

\begin{figure}[t]
    \centering
    \includegraphics[width=0.48\linewidth]{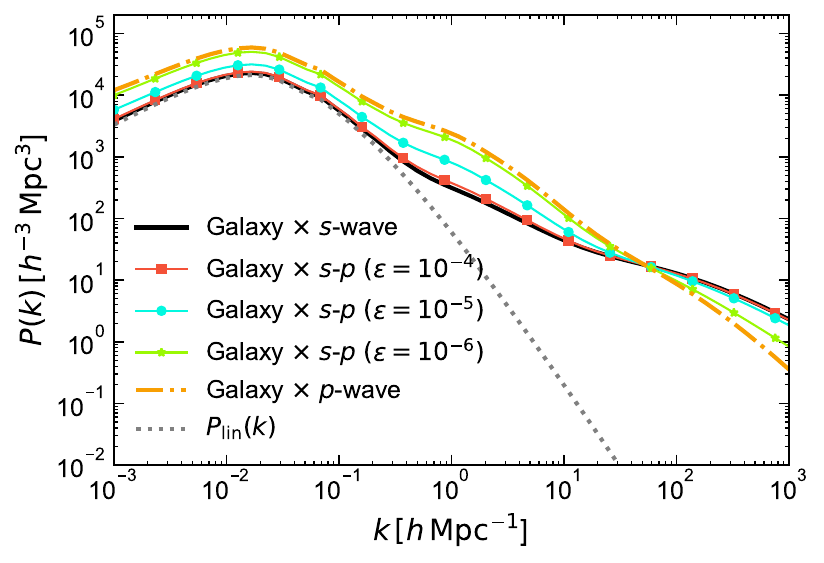}
    \includegraphics[width=0.48\linewidth]{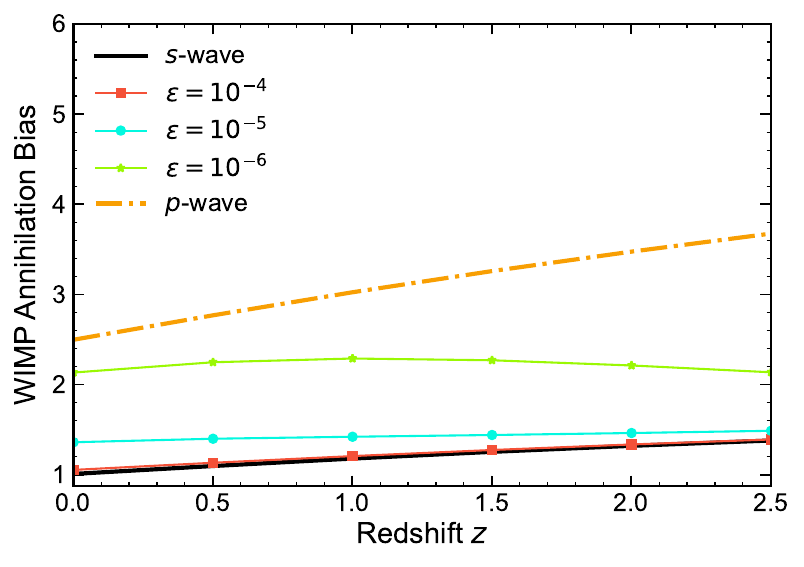}
    \caption{\textit{Left}: Three-dimensional cross-power spectra at
    $z=0.05$ between the WIMP annihilation field and the galaxy
    distribution, for the
    $s$--$p$ admixture scenarios with $\epsilon=10^{-4}$ (red squares),
    $\epsilon=10^{-5}$ (cyan circles), and $\epsilon=10^{-6}$
    (lime-green stars). The $s$-wave and $p$-wave cases are shown
    by the solid black and dash-dotted orange curves, respectively, for
    comparison. \textit{Right}: The corresponding redshift evolution of
    the effective bias $b_{\rm DM}(z)$ for the same set of scenarios.
    The gray dotted line in the left panel shows the linear matter power
    spectrum at $z=0.05$.}
    \label{fig:CPS_bias_spwave}
\end{figure}

\begin{figure}[t]
    \centering
    \includegraphics[width=0.48\linewidth]{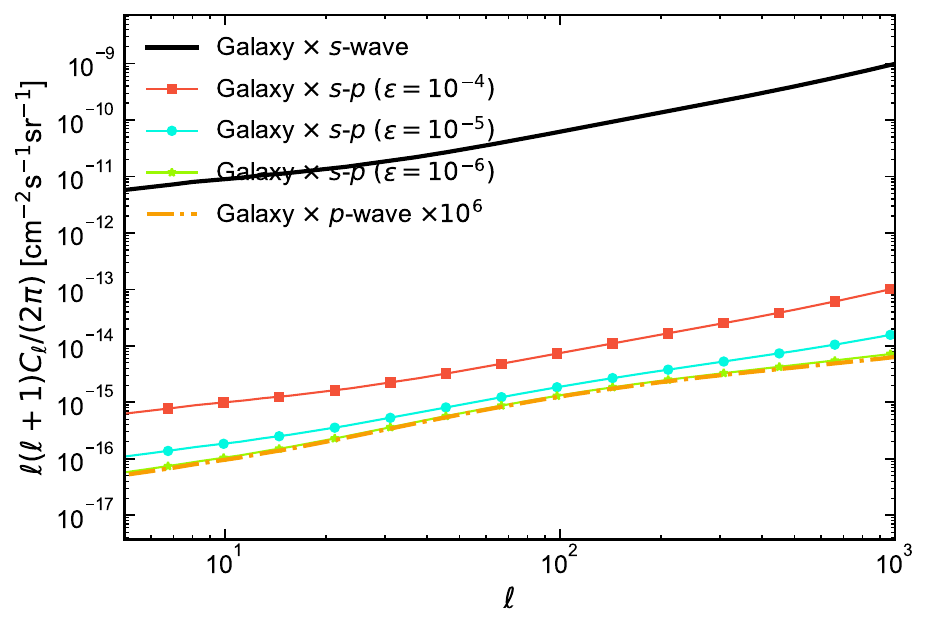}
    \includegraphics[width=0.48\linewidth]{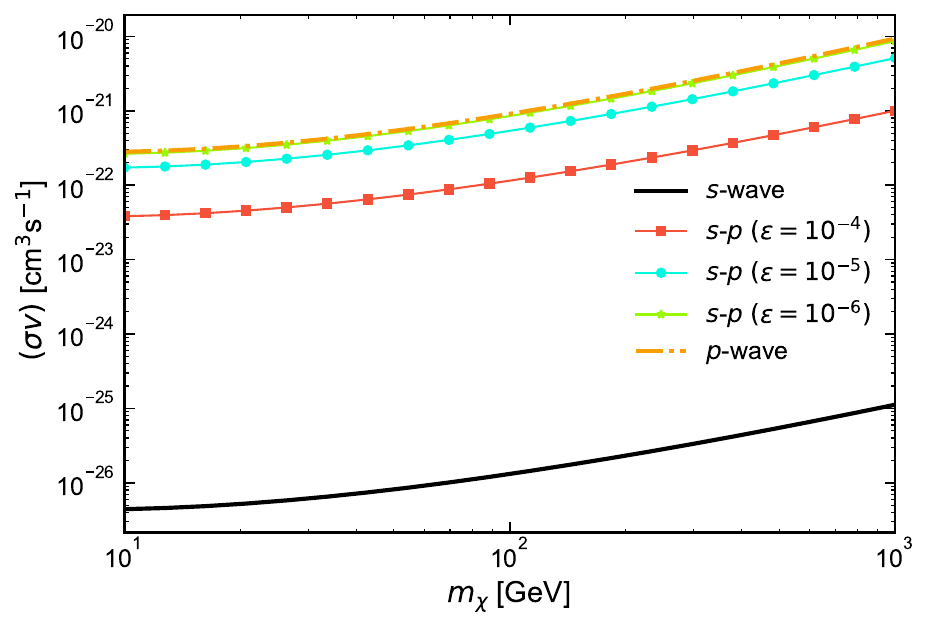}
    \caption{\textit{Left}: Angular cross-power spectra between the
    WIMP annihilation field and the galaxy distribution, integrated over
    $0\leq z\leq 0.1$, in the $(4.8,\,8.3)\,\mathrm{GeV}$ energy bin.
    The $s$--$p$ admixture scenarios with $\epsilon=10^{-4}$ (red
    squares), $\epsilon=10^{-5}$ (cyan circles), and $\epsilon=10^{-6}$
    (lime-green stars) are shown for a benchmark WIMP mass of
    $m_{\rm DM}=100\,\mathrm{GeV}$, annihilation into the $b\bar b$
    channel, and a normalization of
    $(\sigma v)=3\times10^{-26}\,\mathrm{cm^3\,s^{-1}}$. The pure
    $s$-wave and $p$-wave cases are shown by the solid black and
    dash-dotted orange curves, respectively, for comparison.
    \textit{Right}: Projected $95\%$ confidence-level upper limits on
    $(\sigma v)$ as a function of the WIMP mass for the three fixed
    values of $\epsilon=10^{-4}$, $10^{-5}$, and $10^{-6}$, together
    with the corresponding constraints for the pure $s$-wave (solid
    black) and $p$-wave (dash-dotted orange) scenarios. For each fixed
    $\epsilon$, $(\sigma v)$ is treated as the sole free
    parameter.}
    \label{fig:APS_UpperLimits_spwave}
\end{figure}

}

\twocolumngrid


\end{document}